\documentclass[aps,pre,superscriptaddress,showpacs,showkeys,nofootinbib,12pt]{revtex4}
\usepackage{epsfig}
\usepackage{amssymb,amsmath,amsfonts}
\usepackage{mathtools}
\usepackage{graphicx}
\usepackage{color}
\usepackage{url}
\usepackage{xfrac}
\usepackage[letterpaper,margin=1.0in]{geometry}

\newcommand{\numberthis}{\addtocounter{equation}{1}\tag{\theequation}}

\begin{document}

\title{The Growth of Oligarchy in a Yard-Sale Model of Asset Exchange:\\
A Logistic Equation for Wealth Condensation
}
\thanks{\copyright 2016, all rights reserved}

\author{Bruce M. Boghosian}
\affiliation{Department of Mathematics, Tufts University, Medford, Massachusetts 02155, USA}
\author{Adrian Devitt-Lee}
\affiliation{Department of Mathematics, Tufts University, Medford, Massachusetts 02155, USA}
\author{Hongyan Wang}
\affiliation{Department of Mathematics, Tufts University, Medford, Massachusetts 02155, USA}
\date{11 January 2016}
\begin{abstract}
\begin{center}
{\bf Abstract}
\end{center}
The addition of wealth-attained advantage (WAA) to the Yard-Sale Model (YSM) of asset exchange has been demonstrated to induce wealth condensation.  In a model of WAA for which the bias is a continuous function of the wealth difference of the transacting agents, the condensation was shown to arise from a second-order phase transition to a coexistence regime.  In this paper, we present the first analytic time-dependent results for this model, by showing that the condensed wealth obeys a logistic equation in time.
\end{abstract}

%\onecolumn
\maketitle
%\normalsize \vfill

\section{\uppercase{Introduction}}
\label{sec:Introduction}

The scientific study of wealth inequality is motivated by a desire to understand not only the drastic unevenness in the distribution of wealth today, but also its dynamism.  Virtually every metric of wealth inequality is changing in the direction of increased concentration of wealth.  For example, in 2010 Oxfam International noted that 388 individuals in the world had as much wealth as half the human population.  They have been publishing this figure annually, and by early 2016 it had reduced to 62 individuals~\cite{bib:Oxfam2016}.

An important class of wealth distribution models that have been analyzed using mathematical methods of statistical physics are called Asset Exchange Models (AEMs)~\cite{bib:Angle,bib:Hayes}.  These have been used to describe the collective behavior of large economic systems based on simple, idealized microscopic rules.  In a simple AEM, there exists a collection of $N$ agents, each of whom possesses some wealth.  The agents exchange that wealth in pairwise transactions.  There are a variety of models describing these transactions.  Most conserve the total number of agents, $N$, and the total wealth in the system, $W$, though extended versions of these models are capable of accounting for wealth redistribution, changes in agent population, the production and consumption of wealth, and multi-agent transactions.

The AEM examined in this paper is a modification of the basic Yard Sale Model (YSM) of asset exchange~\cite{bib:Chakraborti2002,bib:Hayes}, in which agents exchange wealth solely by means of pairwise transactions.  When two agents enter such a transaction, each has the same probability of winning some amount of wealth from the other, and the amount won is equal to a fraction of the \emph{poorer} agent's wealth.

Following the work of Ispolatov \emph{et al.}~\cite{bib:IspolatovKrapivskyRedner}, Boghosian derived a Boltzmann equation for the basic YSM~\cite{bib:Boghosian1}.  In the limit of small transactions, he showed that the Boltzmann equation reduces to a particular Fokker-Planck (FP) equation, and later demonstrated that this FP equation could be derived much more simply from a stochastic process~\cite{bib:Boghosian2}.  In the absence of any kind of wealth redistribution, Boghosian \emph{et al.}~\cite{bib:Boghosian3} proved that all of the wealth in the system is eventually held by a single agent.  This is due to a subtle but inexorable bias in favor of the wealthy in the rules of the YSM:  Because a fraction of the poorer agent's wealth is traded, the wealthy do not stake as large a fraction of their wealth in any given transaction, and therefore can lose more frequently without risking their status.  This is ultimately due to the multiplicative nature of the transactions on the agents' wealth, as pointed out by Moukarzel~\cite{bib:Moukarzel2007}.

In the above-mentioned works, Boghosian \emph{et al.}~\cite{bib:Boghosian1,bib:Boghosian2,bib:Boghosian3} also investigated the addition of a simple Ornstein-Uhlenbeck-like model of redistribution to the YSM~\cite{bib:OU}.  They demonstrated that it suppressed the tendency of all the wealth to go to a single agent, resulting in a classical distribution, and exhibiting some similarity with empirical forms for wealth distributions due to Pareto~\cite{bib:Pareto} and Gibrat~\cite{bib:Gibrat}.  In later work, however, they demonstrated that the extreme tail of this distribution decays as a gaussian~\cite{bib:Boghosian4}.

The phenomenon of {\it wealth condensation} was first described by Bouchard and M\'{e}zard in 2000, who noted the accumulation of macroscopic levels of weath by a single agent in a simple model of trading and redistribution~\cite{bib:BouchaudMezard2000}.  In 2007, Moukarzel \emph{et al.} investigated wealth-attained advantage (WAA) in the YSM by adding a fixed bias to the probability of winning in any transaction, dependent only on the sign of the wealth differential.  He observed a first-order phase transition to a wealth-condensed state of {\it absolute oligarchy}, in which a single agent held all the wealth~\cite{bib:Moukarzel2007}.

More recently, Boghosian \emph{et al.}~\cite{bib:Boghosian4} introduced a new model for WAA in the YSM, with  bias favoring the wealthier agent proportional to the wealth differential between the two agents, thus approaching zero continuously for transactions between agents of equal wealth.  This model exhibits a second-order phase transition to a state of coexistence between an oligarch and a classical distribution of non-oligarchs.  In that work it was also demonstrated that the above-mentioned gaussian tail was present both below and above criticality, but degenerated to exponential decay precisely at criticality.

While it is perhaps unsurprising that WAA promotes the condensation of wealth, the above observation demonstrates that the way it is introduced can have macroscopic consequences.  In a first-order phase transition, order parameters, such as the Gini coefficient or the fraction of wealth held by the wealthiest agent in this case, are discontinuous functions of the control parameters.  In a second-order phase transition, they exhibit only slope discontinuities.  It seems, therefore, that the continuity or discontinuity of the bias in the microscopic model is directly reflected in the continuity or discontinuity of the macroscopic order parameter.

To be specific, if the coefficient $\tau_\infty$ measures the level of redistribution for the wealthiest agents, and $\zeta$ measures the level of WAA (in a fashion made precise in \cite{bib:Boghosian4}), then criticality was shown to occur at $\zeta=\tau_\infty$, and coexistence for $\zeta > \tau_\infty$.  The fraction of wealth held by the oligarch in the continuum limit was shown to be
\begin{equation}
c_\infty =
\left\{
\begin{array}{ll}
0 & \mbox{if $\zeta\leq\tau_\infty$}\\
1 - \frac{\tau_\infty}{\zeta} & \mbox{if $\zeta > \tau_\infty$}
\end{array}
\right.
\label{eq:cInfty}
\end{equation}
Note that this is a continuous function, with a discontinuous first derivative at the critical point $\zeta=\tau_\infty$, reflective of a second-order phase transition.

Note that all of the above-described observations were made for the steady state situation.  In this paper we quantify the time dependence of the formation of partial oligarchy in the model~\cite{bib:Boghosian4}.  We derive a PDE, valid in the coexistence regime $\zeta > \tau_\infty$, governing the distribution of wealth $p(w,t)$ amongst the non-oligarchs, coupled with an ODE for the fraction of wealth held by the oligarch, $c(t)$.  The latter is the logistic equation
\begin{equation}
c'(t) = c(t)\left[-\tau_\infty + \zeta\left(1-c(t)\right)\right],
\end{equation}
whose long-time limit $c_\infty := \lim_{t\rightarrow\infty} c(t)$ is consistent with Eq.~(\ref{eq:cInfty}) for $\zeta > \tau_\infty$.

In Section~\ref{sec:YSM} we describe the YSM, and the derivation of the FP equation describing its behavior.  In particular, we review the assumptions and methodology of the Kramers-Moyal derivation of the FP equation from a stochastic process, because these assumptions are violated by the singular distributional solutions that we shall be studying.

In Section~\ref{sec:WAAac} we provide a mathematical description for oligarchy as the presence of a singular distribution $\Xi$, correct the Kramers-Moyal derivation of the FP equation, and  present the logistic ODE that describes the wealth of the oligarch.  For reasons discussed in the conclusions, this decouples from the PDE governing the distribution of non-oligarchs.

\section{\uppercase{The Yard Sale model}}
\label{sec:YSM}

In this section we will introduce notation, discuss the interaction between agents in the modified YSM, and review the assumptions and methodology of the Kramers-Moyal derivation of the FP equation from a stochastic process.  While this section follows that of \cite{bib:Boghosian4} closely, this review is necessary because we shall require a weak form of the FP equation in order to accommodate distributional solutions in what follows.

A continuous distribution of wealth can be described by the agent density function (ADF), $P(w,t)$, defined such that the number of agents with wealth $w \in [a,b]$ at time $t$ is given by $\int_a^b P(w,t) dw$. The zeroth and first moments of the ADF correspond to the total number of agents and wealth,
\begin{align*}
N_P &:= \int_{0}^\infty dw\; P(w,t),\\
W_P &:= \int_{0}^\infty dw\; P(w,t) w.
\end{align*}
We will often need the three following partial moments:
\begin{align*}
A_P(w,t) &:= \int_{w}^\infty dx\; {\frac{P(x,t)}{N_P}},\\
L_P(w,t) &:= \int_{0}^w dx\; {\frac{P(x,t)}{N_P}} x,\\
B_P(w,t) &:= \int_{0}^w dx\; {\frac{P(x,t)}{N_P}} {\frac{x^2}{2}}.
\end{align*}
Here $A_P(w)$ denotes the Pareto potential, which is the fraction of agents with wealth at least $w$.  Also, $L_P(w)$ is the Lorenz potential, which denotes the fraction of wealth held by agents with wealth up to $w$.  In the following sections, the expectation of functions over the ADF will be denoted as $\mathcal E_x[f(x)]:= \int_{0}^\infty dx\; {\frac{P(x,t)}{N_P}} f(x)$.

To describe the dynamics of wealth distributions, we use a random walk based on the YSM, in which two agents are randomly chosen to transact, and the magnitude of wealth exchanged is a fraction of the minimum wealth of the two agents.  The winner is determined by a random variable $\eta \in \{-1,+1\}$.  If we focus on a single agent whose wealth transitions from $z$ to $w$ as a result of the transaction, this gives rise to the random walk
\[
w - z = \eta\sqrt{\Delta t} \min(z,x),
\]
where $\sqrt \Delta t$ is a measure of the transactional timescale and $x$ is the wealth of the ``other agent'' with whom the agent in question interacts. 

We now introduce redistribution:  After each transaction, a wealth tax is imposed on each agent, and the total amount collected is redistributed amongst all the agents in the system.  The fraction collected from an agent with wealth $z$ per unit time is denoted by $\tau(z)$, so the total tax collected per unit time is $T_P(t) = \int_0^\infty dz\; P(z,t) \tau(z)z$.  The average amount returned to each agent per unit time is $T_P/N_P$, and the fractional deviation from this average amount for an agent of wealth $z$ will be denoted by $\sigma(z)$, so the amount returned to that agent is $T_P/N_P + \sigma(z)z$.  Hence the net taxation experienced by an agent with wealth $z$ is
\[
\tau(z)z - \left[\frac{T_P}{N_P} + \sigma(z)z\right] = \rho(z)z - \frac{T_P}{N_P},
\]
where we have defined $\rho(z) := \tau(z) - \sigma(z)$.  Note that $\lim_{z\to\infty}\sigma(z)=0$, so $\tau_\infty := \lim_{z\to\infty}\tau(z)=\lim_{z\to\infty}\rho(z) =:\rho_\infty$.
Because the total amount collected must be equal to the total amount redistributed, the expectation value of the deviation from average redistribution must vanish, whence $0 = {\mathcal E}_z[\sigma(z)z]$ and
\begin{equation}
T_P(t) = \int_0^\infty dz\; P(z,t) \rho(z)z.
\label{eq:T}
\end{equation}

For example, if we assume that $\tau(z)$ is a constant, independent of $z$, we find that
\[
T_P(t) = \int_0^\infty dz\; P(z) \tau z = \tau W_P
\]
is also constant.  If we further take $\sigma(z)=0$, so that the redistribution is uniform, we see that the total effective taxation rate is
\[
\tau z - \frac{T_P}{N_P} = \tau z - \frac{\tau W_P}{N_P} = \tau\left(z - \frac{W_P}{N_P}\right).
\]
This redistribution model is reminiscent of the Ornstein-Uhlenbeck process~\cite{bib:OU}, and has been used in  recent studies of the redistributive YSM~\cite{bib:Boghosian4}.

The addition of redistribution determines a new random walk for an individual agent, defined by
\begin{equation}
w - z = -\left(\rho(z)z - \frac{T_P}{N_P}\right)\Delta t + \eta\sqrt{\Delta t}  \min(z,x).
\label{RandomWalk}
\end{equation}
We see that only the difference $\rho(z) := \tau(z) - \sigma(z)$ matters, in terms of which $T_P(t)$ is given by Eq.~\eqref{eq:T}.  In what follows, we shall allow for $\rho$ to be negative, but we will require that it be bounded by a polynomial.

The random variable $\eta$ can be distributed fairly (with mean $\mathcal E[\eta] = 0$), or it can be biased.  The bias  assumed in this paper is that for a model of WAA described in an earlier paper~\cite{bib:Boghosian4}, specifically
\begin{equation}
\mathcal E[\eta] = \zeta {\frac{N_P}{W_P}}\sqrt{\Delta t}(z-x),
\label{eq:WAABias}
\end{equation}
where $\zeta$ is a positive parameter that skews the probability of winning in favor of the wealthier agent.  Thus, in this model, the bias is determined by the difference between the wealth of the two agents.

We will work from Eq.~\eqref{RandomWalk} from the start, and simpler systems can be obtained by letting $\zeta = 0$ or $\tau(w) = 0$. We shall first formally derive a PDE which is implied by the random walk in the limit as the fraction of wealth traded becomes infinitesimal.

We suppose that the probability of the transition $(z,t)\to(w,t+\Delta t)$ is $p_{\Delta t}(z\to w;z,t)$, and that this probability distribution is normalized.  The Chapman-Kolmogorov equation for this random walk is then
\[
P(w,t+\Delta t) = \int_{0}^\infty dz\; P(z,t) p_{\Delta t}(z\to w;z,t)
\]
To derive the corresponding FP equation satisfied by $P$, we must cast the above in weak form.  To this end, we let $\psi$ be an analytic Schwartz function on the domain $[0,\infty)$, and compute its $L^2$ inner product with $\partial P/\partial t$ as follows,
%{\scriptsize
\begin{align*}
%\MoveEqLeft
\left\langle \psi, {\frac{\partial P}{\partial t}}\right\rangle %\\
&= \int\psi(w) {\frac{\partial P(w,t)}{\partial t}} \\
&=\lim_{\Delta t\rightarrow 0} {\frac{1}{\Delta t}}\left [\int dw \,\psi(w) P(w,t+\Delta t) %\right.\\
%&\;\;\;\;\;\;\;\;\;\;\;\;\;\;\;\;\;\;\;\; \left. 
- \int dw\, \psi(w) P(w,t)\right] \\
&=\lim_{\Delta t\rightarrow 0} {\frac{1}{\Delta t}}\int dz \,P(z,t) \int dw \,p_{\Delta t}(z\rightarrow w;z,t)%\\
%&\;\;\;\;\;\;\;\;\;\;\;\;\;\;\;\;\;\;\;\; 
[\psi(w)-\psi(z)]\\
&=\lim_{\Delta t\rightarrow 0} {\frac{1}{\Delta t}}\int dz \,P(z,t) \int dw \,p_{\Delta t}(z\rightarrow w;z,t)%\\
%&\;\;\;\;\;\;\;\;\;\;\;\;\;\;\;\;\;\;\;\;
\sum_{n=1}^\infty {\frac{(w-z)^n}{n!}}{\frac{\partial ^n\psi(z)}{\partial z^n}} \\
&= \sum_{n=1}^\infty{\frac{1}{n!}}\int dz\, M_n(z,t)P(z,t){\frac{\partial ^n\psi(z)}{\partial z^n}} \numberthis \label{eq:GeneralFP}
\end{align*}
%}
where we have defined
\begin{align*}
M_n(z,t)
&:= \lim_{\Delta t\rightarrow 0} {\frac{1}{\Delta t}}\int dw \,p_{\Delta t}(z\rightarrow w;z,t) (w-z)^n \\
&= \mathcal E_{x,\eta}\left[{\frac{(w-z)^n}{\Delta t}}\right].
\end{align*}
The above result may be written
\begin{equation}
\left\langle \psi, {\frac{\partial P}{\partial t}}\right\rangle
= \sum_{n=1}^\infty{\frac{1}{n!}}\left\langle M_n(z,t)P(z,t),{\frac{\partial ^n\psi(z)}{\partial z^n}}\right\rangle.
\end{equation}
whence, using integration by parts to revert to the strong form, we obtain
\begin{equation}
{\frac{\partial P}{\partial t}}
= \sum_{n=1}^\infty{\frac{(-1)^n}{n!}}\partial_{z}^n\left[M_n(z,t)P(z,t)\right],
\label{FP}
\end{equation}
and thus the agent density function satisfies a PDE that is determined by the moments of the random walk on an infinitesimal timescale.

\subsection{Fokker-Planck equation}
\label{ssec:FP}

Recalling Eqs.~(\ref{RandomWalk}) and \eqref{eq:WAABias}, we can compute $M_1$,
%{\scriptsize
\begin{align*}
%\MoveEqLeft 
M_1(z,t)
&= \lim_{t\rightarrow 0} \mathcal E_{\eta,x}\left[{\frac{(w-z)}{\Delta t}}\right] \\
&= \mathcal E_x\left[\frac{T_{P}(t)}{N_P}-z\rho(z) + \zeta {\frac{N_P}{W_P}}(z-x)\min(w,x)\right] \\
&= \frac{T_{P}(t)}{N_P}-z\rho(z) - \zeta \left[2{\frac{N_P}{W_P}}B_P(z,t) -2zL_P(z,t)
- z^2{\frac{N_P}{W_P}}A_P(z,t) + z\right].
\numberthis\label{NormalM1}
\end{align*}
%}
Next, because $\eta\in\{-1,+1\}$, we have $\mathcal E_\eta[\eta^2] = 1$, which allows us to compute $M_2$,
%{\scriptsize
\begin{align*}
%\MoveEqLeft
M_2(z,t)
&=\lim_{t\rightarrow 0} \mathcal E_{\eta,x}\left[{\frac{(w-z)^2}{\Delta t}}\right] \\
&= \lim_{t\rightarrow 0}\mathcal E_{\eta,x}\left[\Delta t\left(\frac{T_{P}(t)}{N_P}-z\rho(z)\right)^2\right. \\
&\;\;\;\;\;\;\;\;\;\;\;\;\;\;\;\;\;\;\;\;\left.+ \left(\frac{T_{P}(t)}{N_P}-z\rho(z)\right)\eta\min(z,x)(\Delta t)^{1/2}+ \eta^2\min(z,x)^2\right] \\
&= \mathcal E_{\eta,x}\left[\eta^2\min(z,x)^2\right] \\
&= \int_{0}^\infty dx\; {\frac{P(x,t)}{N_P}}\min(z,x)^2 \\
& = 2B_P(z,t) + z^2A_P(z,t).
\numberthis \label{NormalM2}
\end{align*}
%}
Finally, we note that higher powers of $\eta$ have expectations $\mathcal E_\eta[\eta^k] \leq 1$ for $k \geq 3$.  It follows that, because each term in the expansion of the above equation includes some positive power of $\sqrt{\Delta t}$,  all of the moments approach zero for $k\geq 3$, 
%{\scriptsize
\begin{align*}
M_k(z,t) &= \lim_{t\rightarrow 0} {\frac{1}{\Delta t}}\mathcal E_{\eta,x}\left[\left(\Delta t\left[\frac{T_{P}(t)}{N_P}-z\rho(z)\right] + \eta\sqrt{\Delta t}\min(w,x)\right)^k\right]%\\
%&
= 0.
\numberthis \label{NormalMk}
\end{align*}
%}

Substituting Eqs.~(\ref{NormalM1}), \eqref{NormalM2} and \eqref{NormalMk} into Eq.~\eqref{FP}, we find that our wealth distribution obeys the following quadratically nonlinear, integrodifferential FP equation,
%{\small
\begin{align*}
{\frac{\partial P}{\partial t}} &+ {\frac{\partial}{\partial w}}\left[\left(\frac{T_{P}(t)}{N_P}-w\rho(w)\right)P \right]\\
&= {\frac{\partial^2}{\partial w^2}}\left[\left(B_P + {\frac{w^2}{2}}A_P\right)P\right]\\
& \phantom{=}+ {\frac{\partial}{\partial w}}\left\{\zeta \left({\frac{N_P}{W_P}}2B_P -2wL_P - w^2{\frac{N_P}{W_P}}A_P + w\right)P\right\}.
\numberthis
\label{eq:GeneralYSM}
\end{align*}
%}

This system conserves wealth and agents, as was shown in the appendix of the paper where it was first derived~\cite{bib:Boghosian4}.  In the case where there is no redistribution and no WAA, we have $\tau=\zeta = 0$, so the agent density function satisfies
\begin{equation}
{\frac{\partial P}{\partial t}} = {\frac{\partial^2}{\partial w^2}}\left[\left(B_P(w,t) + {\frac{w^2}{2}}A_P(w,t)\right)P\right].\label{SimpleYSM}
\end{equation}

\subsection{Oligarchy as a distributional solution}
\label{ssec:Oligarchy}

The steady-state solution to Eq.~\eqref{eq:GeneralYSM} may involve the condensation of a finite fraction of the system's wealth into the hands of a vanishingly small number of agents.  Indeed, in the absence of redistribution, as in Eq.~\eqref{SimpleYSM}, all the wealth will condense in this manner.  To describe this, we must extend our function space to include certain singular distributions.

If our system were discrete, then complete concentration of wealth would be described by $P(w) = (N_P-1)\delta(w) + \delta(w-W_P)$, where $N_P-1$ agents have no wealth, and a single agent holds all of the wealth. In the continuum limit, however, the number of agents need not be an integer.  Thus wealth condensation could continue indefinitely, with a ``half an agent'' holding twice the wealth of the system, described by the distribution $P(w) = (N_P - \sfrac{1}{2})\delta(w) + {\sfrac{1}{2}}\delta(w-2W_P)$.  More generally, we can take the distribution to be $P(w) = (N_P-\epsilon)\delta(w) + \epsilon \delta(w - W_P/\epsilon)$, where $\epsilon$ is an arbitrarily small positive number.  Replacing $\epsilon$ by $W_P\epsilon$ and passing to the limit as $\epsilon\rightarrow 0$ yields
\begin{equation}
P(w) = N_P\delta(w) + W_P\Xi(w),
\end{equation}
where we have defined $\Xi(w) = \lim_{\epsilon\rightarrow 0}\epsilon\delta(w - 1/\epsilon)$.  To be more precise, $\Xi$ should be defined as the singular distribution whose action on a test function $\phi$ is
\begin{align*}
\langle \Xi, \phi\rangle
&= \lim_{\epsilon\rightarrow 0} \epsilon \left\langle\delta\left(w-{\frac{1}{\epsilon}}\right),\phi\right\rangle\\
&= \lim_{\epsilon\rightarrow 0} \epsilon\;\phi\left(\frac{1}{\epsilon}\right)\\
&= \lim_{s\rightarrow\infty}\frac{\phi(s)}{s}.
\numberthis\label{DefXi}
\end{align*}
Here the space of test functions to which $\phi$ belongs is
%{\small
\[
\partial^{-2}\mathcal S_0
= \{\phi(w) = \psi(w) + \gamma + \mu w: \psi(w)\in\mathcal S([0,\infty)), \gamma,\mu\in \mathbb R\},
%\label{Def2test}
\]
%}
where $\mathcal{S}([0,\infty)$ denotes Schwartz functions on $[0,\infty)$.

\section{\uppercase{WAA above criticality}}
\label{sec:WAAac}

In this section, we use the definition of $\Xi$ provided by Eq.~\eqref{DefXi}.  In earlier work~\cite{bib:Boghosian4}, the FP equation, Eq.~\eqref{eq:GeneralYSM}, and its steady-state asymptotic behavior were discussed in detail.  A second-order phase transition was observed at the critical point $\zeta = \tau_\infty$, and coexistence was observed between the singular distribution $\Xi$ and a classical distribution for $\zeta > \tau_\infty$.  This may be thought of as an oligarch in coexistence with a population of non-oligarchs.  The presence of the singular distribution $\Xi$, however, violates the assumptions in the derivation of the FP equation, as our presentation leading to Eq.~\eqref{FP} assumed an analytic Schwartz function in the weak form.

Recall that the mathematical phenomenon that we describe here as ``oligarch'' may be modeled as the wealth held by the wealthiest fraction $\epsilon$ of agents, as $\epsilon\rightarrow 0$.  Assume that there exists a wealth distribution which is a steady state of the random process, Eq.~\eqref{RandomWalk}, in the small-transaction limit.  Moreover, suppose that this distribution may be written as $P(w) = p(w) + c(t)W_P\Xi(w)$, where $p(w)$ is the classical agent density function for the non-oligarchs.  This is a system in which the wealthiest infinitesimal of agents hold a fraction $c(t)$ of the total wealth of the system at time $t$.  We consider two sets of agents in the random walk:  There are the ``normal'' agents, corresponding to the classical distribution $p(w)$.  Then there is what we have been calling the ``oligarch,'' considered to be the wealthiest fraction $\epsilon$ of the distribution $P(w)$, considered as a single unit.

Now suppose that we have an oligarch, and $\phi = \psi + \gamma+ \mu w$, where $\psi$ is an analytic Schwartz function.  Then the development that led to Eq.~\eqref{eq:GeneralFP} becomes:
%{\scriptsize
\begin{align*}
\left\langle{\frac{\partial P}{\partial t}},\phi\right\rangle &= \sum_{n=1}^\infty{\frac{1}{n!}}\int M_n(z,t)P(z,t){\frac{\partial ^n\phi(z)}{\partial z^n}} dz \\
&= \sum_{n=1}^\infty{\frac{1}{n!}}\int\left[M_n(z,t)P(z,t)\right]\partial_{z}^n\psi + \mu M_1(z,t)P(z,t)dz \\
&= \left(\sum_{n=1}^\infty{\frac{1}{n!}}\int\partial_{z}^n\left[M_n(z,t)P(z,t)\right]\psi\, dz\right) + \mu N_P\mathcal E_z[M_1(z,t)].
\numberthis \label{WeakFP}
\end{align*}
%}
In strong form we therefore have
%{\small
\begin{equation*}
{\frac{\partial P}{\partial t}}=  \left(\sum_{n=1}^\infty{\frac{(-1)^n}{n!}}\partial_{z}^n\left[M_n(z,t)P(z,t)\right]\right)+ \mathcal E_z[M_1(z,t)]N_P\Xi,
\label{XiFP}
\end{equation*}
%}
where the $\Xi$ term is from $\langle M_1(z,t)P(z,t),\mu\rangle = \mu \langle P(z,t),M_1(z,t)\rangle = \langle \Xi,\phi\rangle \cdot N_P\mathcal E_z[M_1(z,t)]$.  This implies two things: the strong form of the FP equation may not make sense in the coexistence regime, and $\Xi$ is likely to depend only on the $M_1$ term, which corresponds to drift in the FP equation.

To carry out the full analysis, we must recalculate the coefficients $M_n(z,t)$ for three cases:  (i) a normal agent interacting with another normal agent, (ii) a normal agent interacting with the oligarch, and (iii) the oligarch interacting with a normal agent.  The first two factors would describe $p(w,t)$, and the last would determine $c(t)$, the fraction of wealth held by the oligarch.  In this paper, we restrict our attention to the wealth of the oligarch only, and we relegate the derivation of the PDE governing the non-oligarchical portion of the distribution to future work.

Denote the first coefficient $M_1$ of the wealthiest $\epsilon$ of agents by $M_1^\epsilon$. Instead of using the full machinery of the FP derivation, we note
%{\scriptsize
\begin{align*}
M_1^\epsilon(z,t) &= \lim_{t\rightarrow 0}\mathcal E_{\eta,x}\left[{\frac{c(t+\Delta t)W_P/\epsilon - c(t)W_P/\epsilon}{\Delta t}}\right] = c'(t) {\frac{W_P}{\epsilon}} \\
&= T_{P}(t) - {c(t)\frac{W_P}{\epsilon}}\tau\left({\frac{c(t)W_P}{\epsilon}}\right) + \mathcal E_{x}\left[\zeta{\frac{N_P}{W_P}}\left({\frac{c(t)W_P}{\epsilon}} - x\right) x\right] \\
&=  T_{P}(t) - (c(t)W_P/\epsilon)\tau({c(t)W_P/\epsilon}) \\
&\phantom{=} + \zeta{\frac{1}{W_P}}[c(t)W_P^2(1-c(t))/\epsilon - 2N_PB_p(c(t)W_P/\epsilon)],
\end{align*}
%}
where the integrals for the expectation are between $x=0$ and $x= cW_P/\epsilon$, so that they do not include the oligarch.  We will assume that $p$ decays like an exponential or a gaussian, as was shown in earlier work~\cite{bib:Boghosian4}.  Under this assumption, the second moment of $p$ is finite, so when we take the limit as $\epsilon \rightarrow 0$, we have
\begin{equation}
\boxed{c'(t) = c(t)[-\tau_\infty + \zeta(1-c(t))],}
\label{eq:logisticC}
\end{equation}
where we have written $\tau_\infty := \lim_{\epsilon\to0}\tau(1/\epsilon)$.  This is a logistic equation for $c(t)$ with solution
\begin{equation}
c(t) = (1-\tau_\infty/\zeta)\frac{c_0e^{(\zeta-\tau_\infty)t}}{c_0(e^{(\zeta-\tau_\infty)t}-1) + 1-\tau_\infty/\zeta}.
\label{OligDynam}
\end{equation}
Eq.~(\ref{eq:logisticC}) is the principal result of this paper.  It indicates that the wealth of the oligarch obeys a logistic equation, independent of the evolution of the classical portion of the wealth distribution.

\begin{figure}[!h]
  \vspace{-0.2cm}
  \centering
   {\epsfig{file = 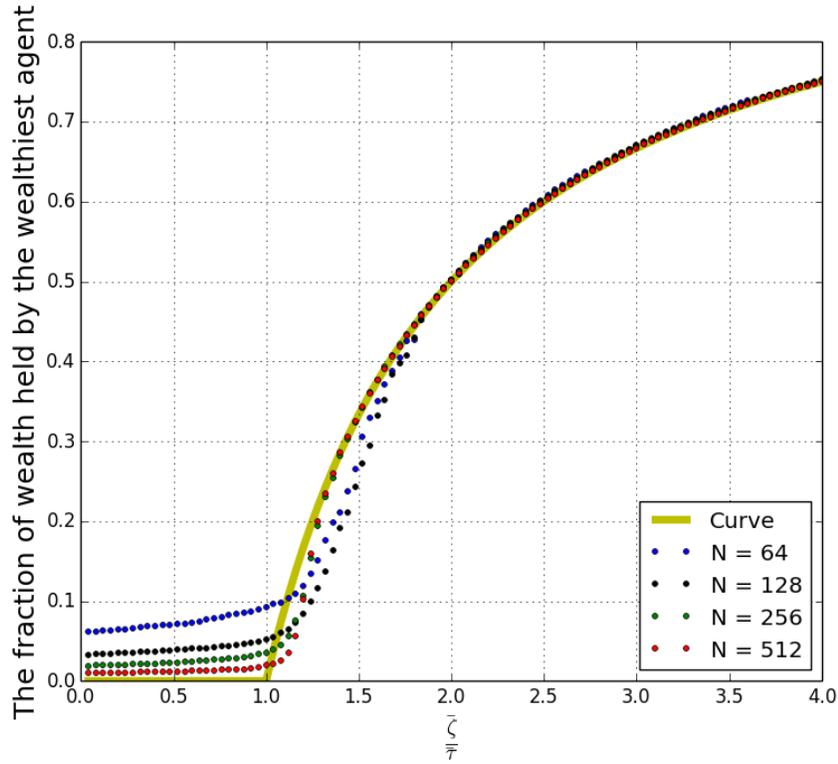, width = 5.0in}}
  \caption{Monte Carlo simulation of the wealth held by the wealthiest agent in simulations of different sizes, with various values of constant $\tau$ and $\sigma$.  As the number of agents grows, this approaches the theoretical result $c = 1-\tau_\infty/\zeta$.}
  \label{fig:MC}
  \vspace{-0.1cm}
\end{figure}

Equation~\eqref{eq:logisticC} indicates that there are two asymptotic solutions in time, one at $c = 0$, and another at $c = 1-\tau_\infty/\zeta$.  Since ${\frac{d}{dc}}\left[c((1-c)\zeta-\tau_\infty)\right]_{c=0} = (\zeta - \tau_\infty)$, the absence of an oligarch is stable if $\zeta < \tau_\infty$, and is unstable otherwise.  Similarly, the presence of an oligarch is stable if $\zeta > \tau_\infty$, and is unstable otherwise~\footnote{Note that we never have a stable oligarch with negative wealth; if $c = 1-\tau_\infty/\zeta < 0$, then $\zeta < \tau_\infty$.}  Furthermore, since the presence of an oligarch implies that there is finite taxation on the wealthiest agent, $\tau_\infty$ must exist.  Monte Carlo simulations confirm the stable steady state of the oligarch both above and below criticality: The wealth held by the wealthiest agent, plotted in Fig.~\ref{fig:MC}, is of the same order as that of all the other agents, and is approximately $c = 1-\tau_\infty/\zeta$ above criticality.

\subsection{Gini coefficients above criticality}
\label{ssec:Gini}

\begin{figure}[!h]
  \vspace{-0.2cm}
  \centering
   {\epsfig{file = 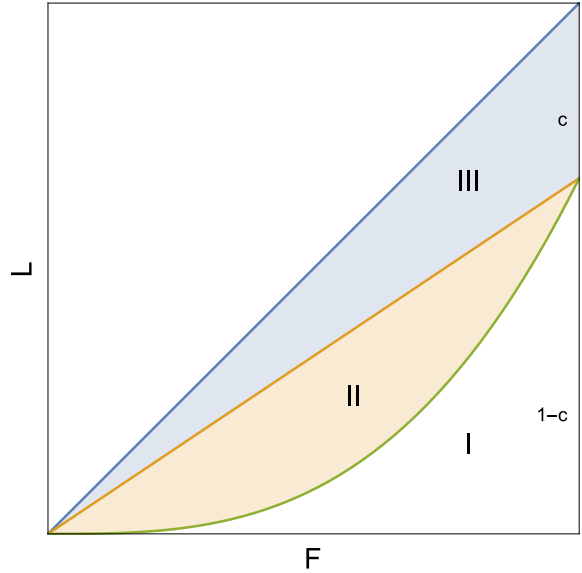, width = 3.0in}}
  \caption{Lorenz plot with partial wealth condensation.  A fraction $c$ of the wealth has condensed, and $1-c$ is distributed classically.}
  \label{fig:Lorenz}
  \vspace{-0.1cm}
\end{figure}

Finally, we wish to define the notion of Gini coefficient in the coexistence region above criticality.  In particular, we wish to describe the Gini coefficient of the entire system in terms of that of the classical system $p$, excluding the oligarch.  The phenomenon of oligarchy is evidenced by a Lorenz curve that does not reach the point (1,1), as illustrated in Fig.~\ref{fig:Lorenz}, where we have labelled three distinct regions below the diagonal.  We shall use the labels ``I'', ``II'' and ``III'' to denote the areas of these regions.
Since the fraction $c$ of the wealth of the system is held by the infinitesimal agent described by the oligarch, the Lorenz curve will reach the point (1,1-c). So if we consider the Gini coefficient of the system $p$, then we know that:
\begin{equation}
G_P = {\frac{II + III}{I + II + III}}, \qquad G_p = {\frac{II}{I + II}} \label{GiniSolve}
\end{equation}
Clearly $I + II + III = 1/2$, $I + II = (1-c)/2$, and $III = c/2$. Using the supercritical value of $c$, this immediately implies that
\begin{align}
G_P &= +1 + {\frac{\tau_\infty}{\zeta}}(1 - G_p)
\label{GiniP} \\
G_p &= -1 + {\frac{\zeta}{\tau_\infty}}(1 - G_P).
\label{Ginip}
\end{align}
These equations are straightforward relations between the versions of the Gini coefficient defined with and without the presence of the oligarch.
 
\section{\uppercase{Discussion and conclusions}}
\label{sec:Conclusions}

Coexistence between wealth-condensed and normal distributions in steady-state solutions of the YSM was first noted in earlier work~\cite{bib:Boghosian4}.  This paper provides the first exact analytic result for the \emph{time-dependent} behavior of that model.  In particular, it demonstrates that the fraction of wealth held by the oligarch above criticality obeys a logistic equation, Eq.~\eqref{eq:logisticC}.  This equation is remarkable in that it is completely decoupled from the classical part of the distribution, $p(w)$, and may be solved exactly.

The reason that the equation for $c(t)$ decouples from that for the classical part of the distribution can be understood by noting that the oligarch always wins in transactional exchanges with non-oligarchs.  From the point of view of the oligarch, the remainder of the distribution might as well be aggregated into a single agent with wealth $W_p = (1-c)W_P$ who, obligingly, always loses in any transaction with the oligarch.  The oligarch's ability to gain wealth from the non-oligarchical portion of the distribution is therefore limited only by the rate at which he/she transacts with non-oligarchs, as compared to the rate at which he/she is taxed.  This is why the steady-state fraction of wealth held by the oligarch depends only on the ratio of tax rate to WAA rate, $\tau_\infty/\zeta$.

From a macroeconomic perspective, it is well known that most real-world oligarchs worry much less than the rest of the population about individual transactions; beyond a certain point, many do not even know where their own money is invested.  By contrast, they worry deeply about how taxation and redistribution affect their fortunes, and they expend significant effort to lobby against progressive taxation, capital gains taxes and inheritance taxes.  We believe that the asymptotic analysis of the YSM may provide a way to understand these priorities.

%\section*{Bibliography}

\end{document}